\documentclass{appolb}
\usepackage{hyperref}
\usepackage{epsfig}
\usepackage{amsmath}%for subequations
\usepackage{amssymb}%for symbols
\usepackage{subfigure}%for subnumbering inside a figure
\def\alf{\alpha_}
\newcommand{\veps}{\varepsilon}

\begin{document}

\title{
%%%%%%%%%%%%%%%%%%
\vspace{-20mm}
\begin{flushright} \bf IFJPAN-IV-2009-2\\ \end{flushright}
\vspace{5mm}
%%%%%%%%%%%%%%%%%%
Non-abelian infra-red cancellations in the unintegrated \uppercase{NLO} kernel.%
\thanks
{This work is supported by the EU grant MRTN-CT-2006-035505, 
and by the Polish Ministry of Science and Higher Education grant 
  No.\ 153/6.PR UE/2007/7.
\hfill \\
  Presented at the
{\em Cracow Epiphany Conference on hadron interactions at the dawn of the LHC}, January 5-7, 2009} 
}%
\author{Magdalena Slawinska and Aleksander Kusina
\address{Institute of Nuclear Physics PAN,\\
ul. Radzikowskiego 152, 31-342 Krak\'ow, Poland }}
\maketitle

\begin{abstract}
{\em Abstract:}
We investigate the infrared singularity structure of Feynman diagrams entering
the next-to-leading-order (NLO) DGLAP kernel (non-singlet).
We examine cancellations between diagrams for two gluon emission
contributing to NLO kernels.
We observe the crucial role of color coherence effects in cancellations of 
infra-red singularities.
Numerical calculations are explained using analytical formulas
for the singular contributions.
\vspace{3mm}
\centerline{\em Submitted To Acta Physica Polonica B}
\end{abstract}


\vspace{5mm}
%%%%%%%%%%%%%%%%%%%%%%%%%%%
\begin{flushleft}
\bf IFJPAN-IV-2009-2\\
\end{flushleft}
%%%%%%%%%%%%%%%%%%%%%%%%%%%

\newpage
%%%%%%%%%%%%%%%%%%%%%%%%%%%%%%%%%%%%%%%%%%%%%%%%%%%%%%%%%%%%%%%%%%%%%%%
%%%%%%%%%%%%%%%%%%%%%%%%%%%%%%%%%%%%%%%%%%%%%%%%%%%%%%%%%%%%%%%%%%%%%%%
\section{Introduction}
%%%%%%%%%%%%%%%%%%%%%%%%%%%%%%%%%%%%%%%%%%%%%%%%%%%%%%%%%%%%%%%%%%%%%%%
This study is part of the effort with the aim
of constructing fully-exclusive (unintegrated) kernels for DGLAP~\cite{DGLAP}
evolution in QCD at the complete NLO level.
More details on this project can be found in ref.~\cite{ifjpan-iv-09-3}.
Let us only mention that construction of the exclusive NLO DGLAP
kernels in ref.~\cite{ifjpan-iv-09-3}
is done following Curci-Furmanski-Petronzio (CFP) scheme
\cite{Curci:1980uw} and we adopt this scheme also in our study.
In short the CFP scheme uses axial gauge
and dimensional regularization ($\overline{MS}$) and generalizes
collinear factorization developed in ref.~\cite{Ellis:1978sf}.
The two particle-irreducible (2PI) evolution kernels $K_0$
\begin{equation}
M = C_0(1+K_0 + K_0^2+ ...) = C_0 \Gamma_0
\end{equation}
are contracted with the coefficient functions $C_0$, which are infra-red finite.
All infra-red collinear divergences
are encapsulated in $\Gamma_0$, which denotes the sum over
kernels $K_0$.
The DGLAP NLO kernel is then extracted according to the scheme~\cite{Curci:1980uw}
as a single pole in $\Gamma_0$.
More details can be found in ref.~\cite{Curci:1980uw} and in
ref.~\cite{ifjpan-iv-09-3} of these proceedings.

As it is well known individual Feynman diagrams are not gauge-invariant
and sizeable cancellations occur among them.
In particular some graphs contributing to NLO DGLAP kernels may
contain artificial infra-red singularities,
which cancel in a bigger subset of diagrams
and do not appear in the final results.
Our aim is to analyze in detail such cancellations at the level
of the exclusive distributions, before the phase space integration,
for the diagrams contributing to NLO DGLAP kernels in the CFP scheme.

\begin{figure}
\begin{centering}
\subfigure[]{
\epsfig{file=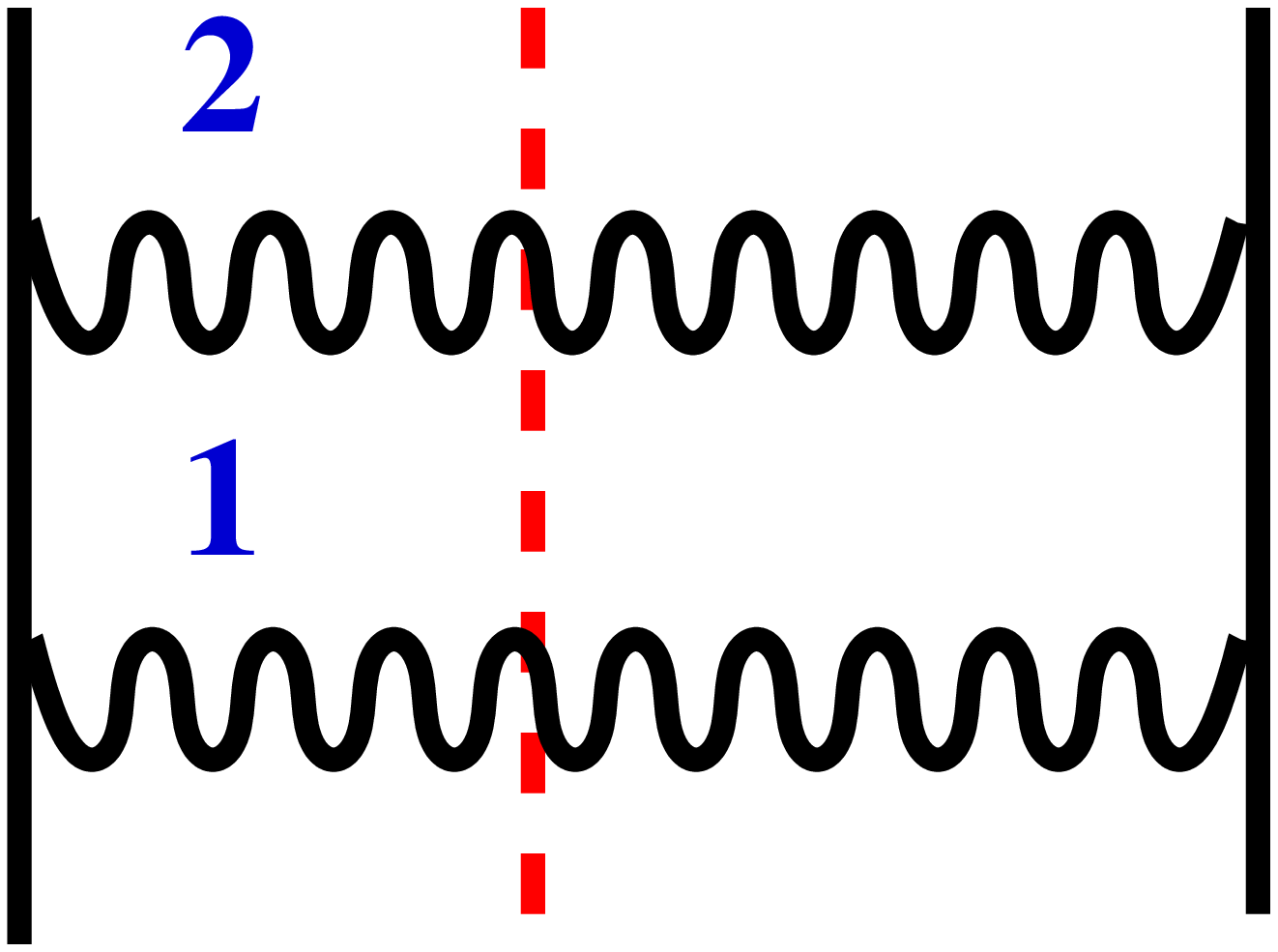, height = 20mm}
  \label{fig:laddera}}
\subfigure[]{
 \epsfig{file=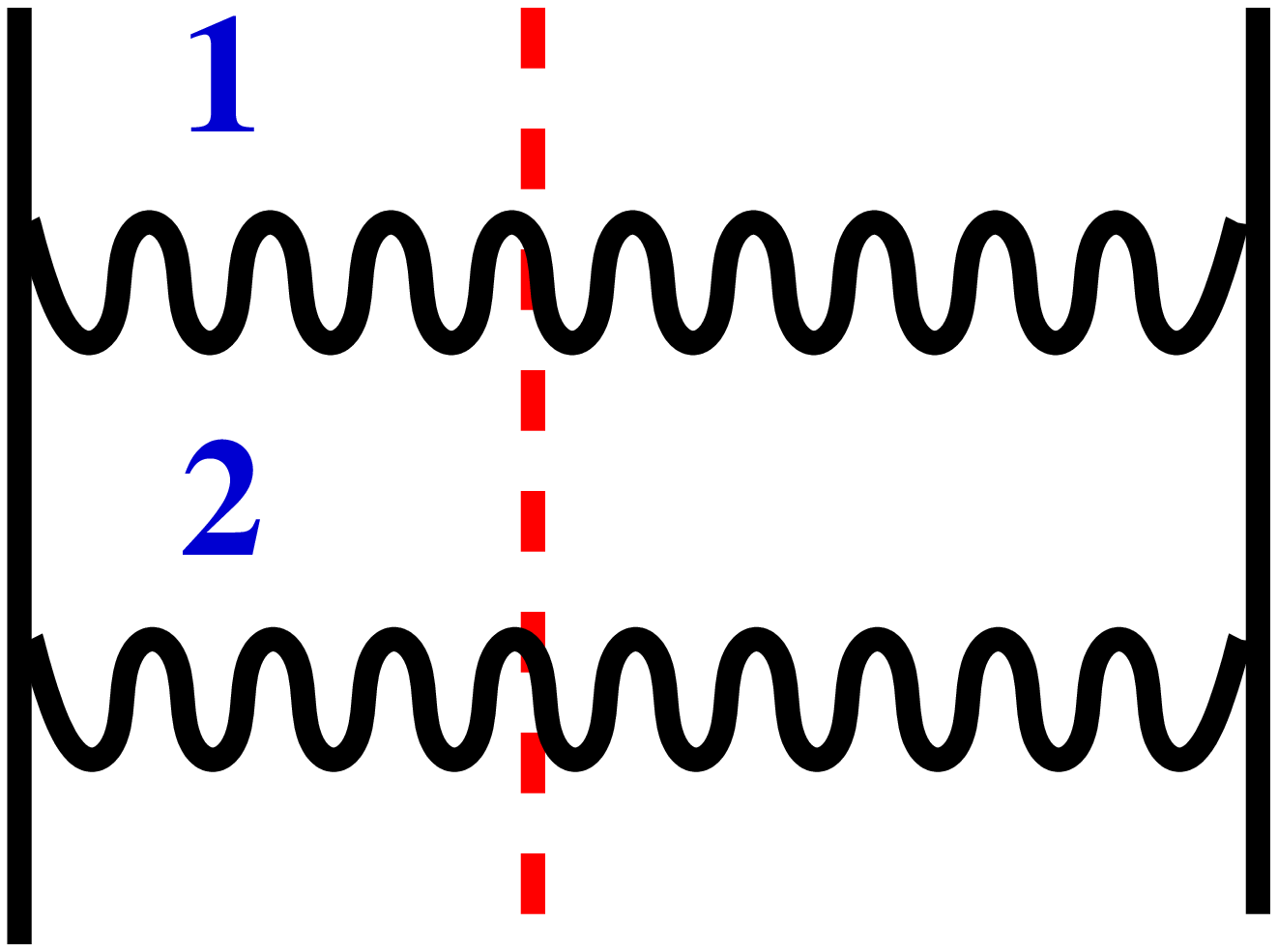, height = 20mm}
 \label{fig:ladderb}}
\subfigure[]{
 \epsfig{file=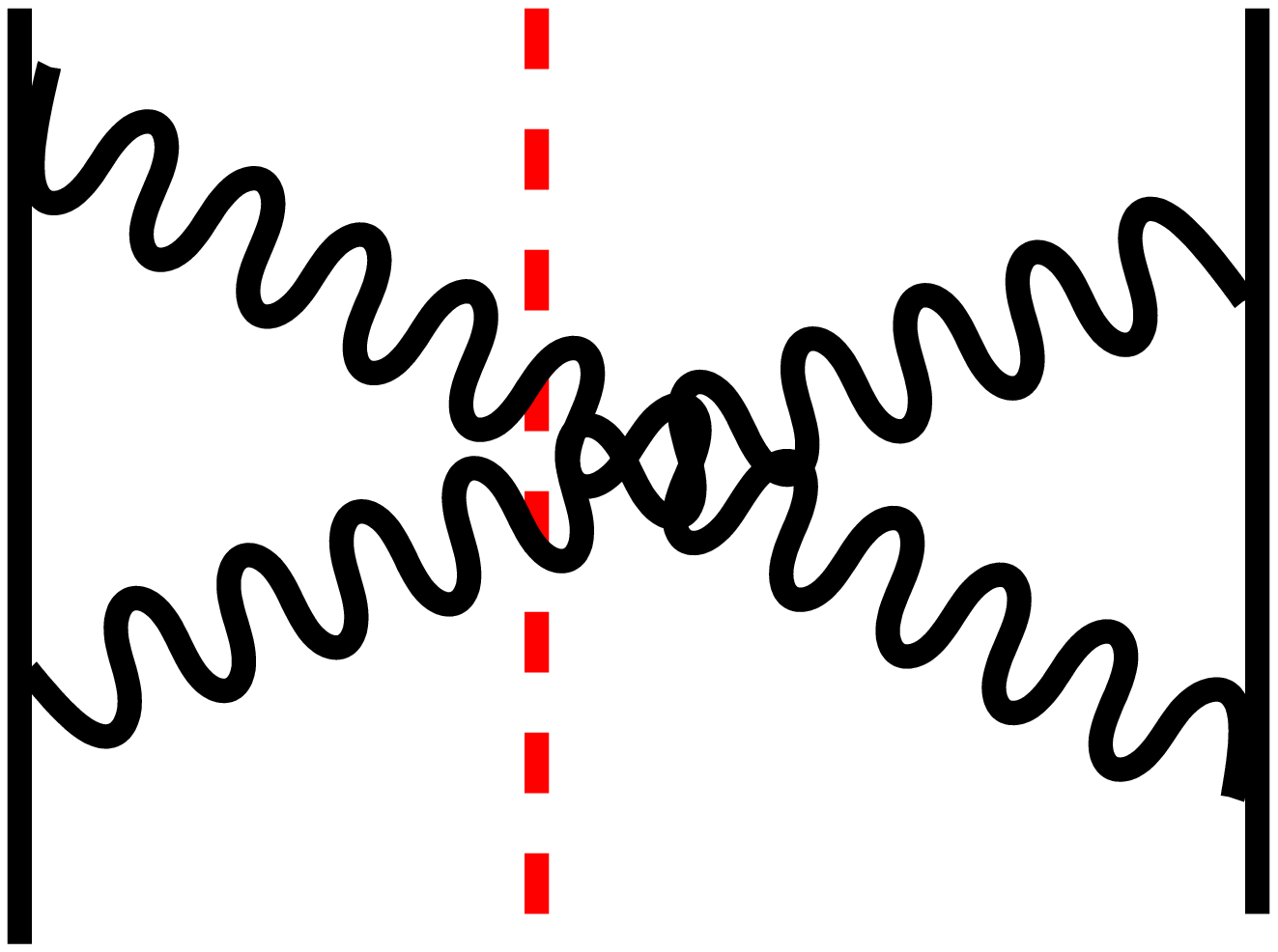,  height = 20mm}
 \label{fig:ladderc}}
\\
\subfigure[]{
\epsfig{file=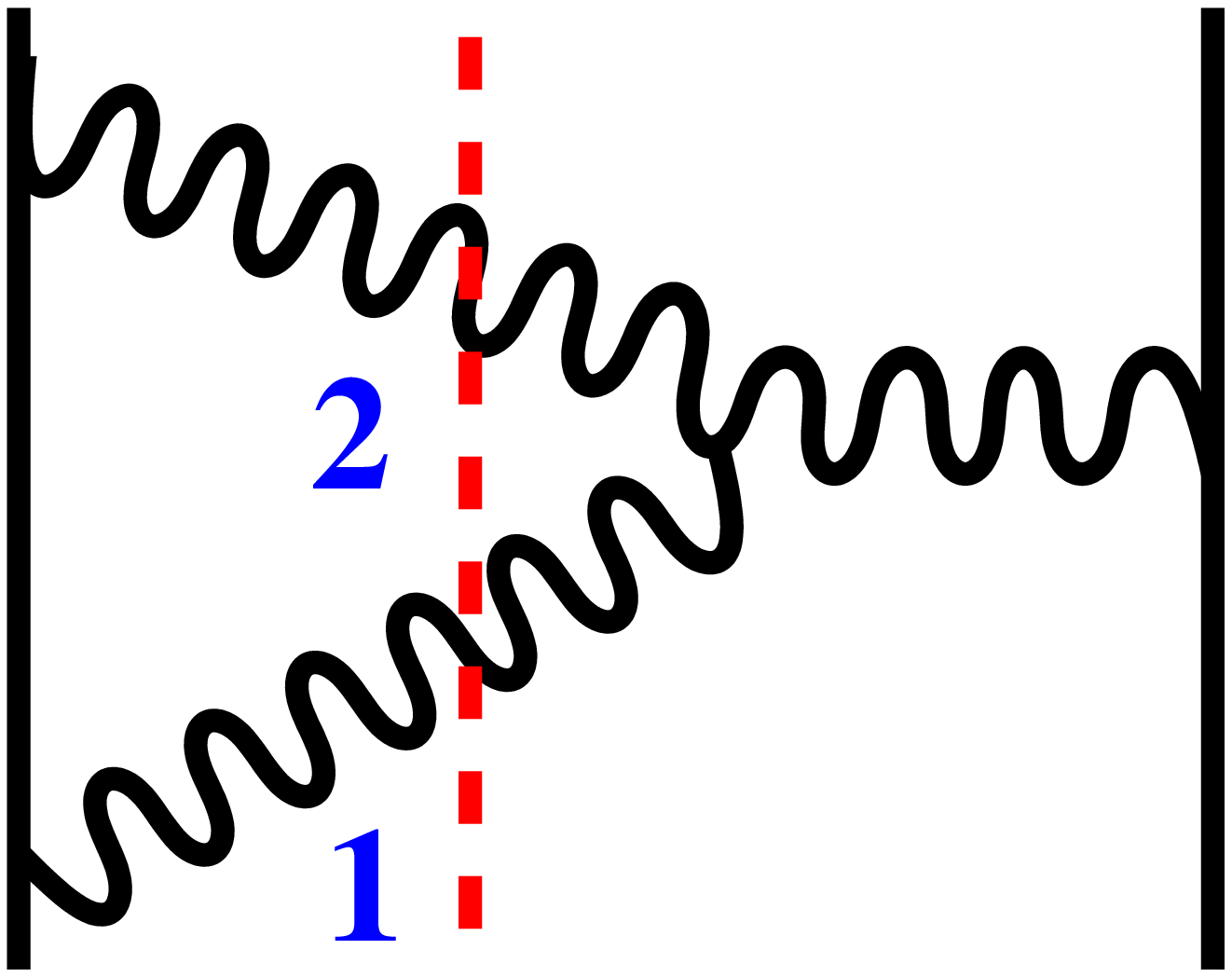, height = 20mm}
  \label{fig:nonabelb}}
\subfigure[]{
\epsfig{file=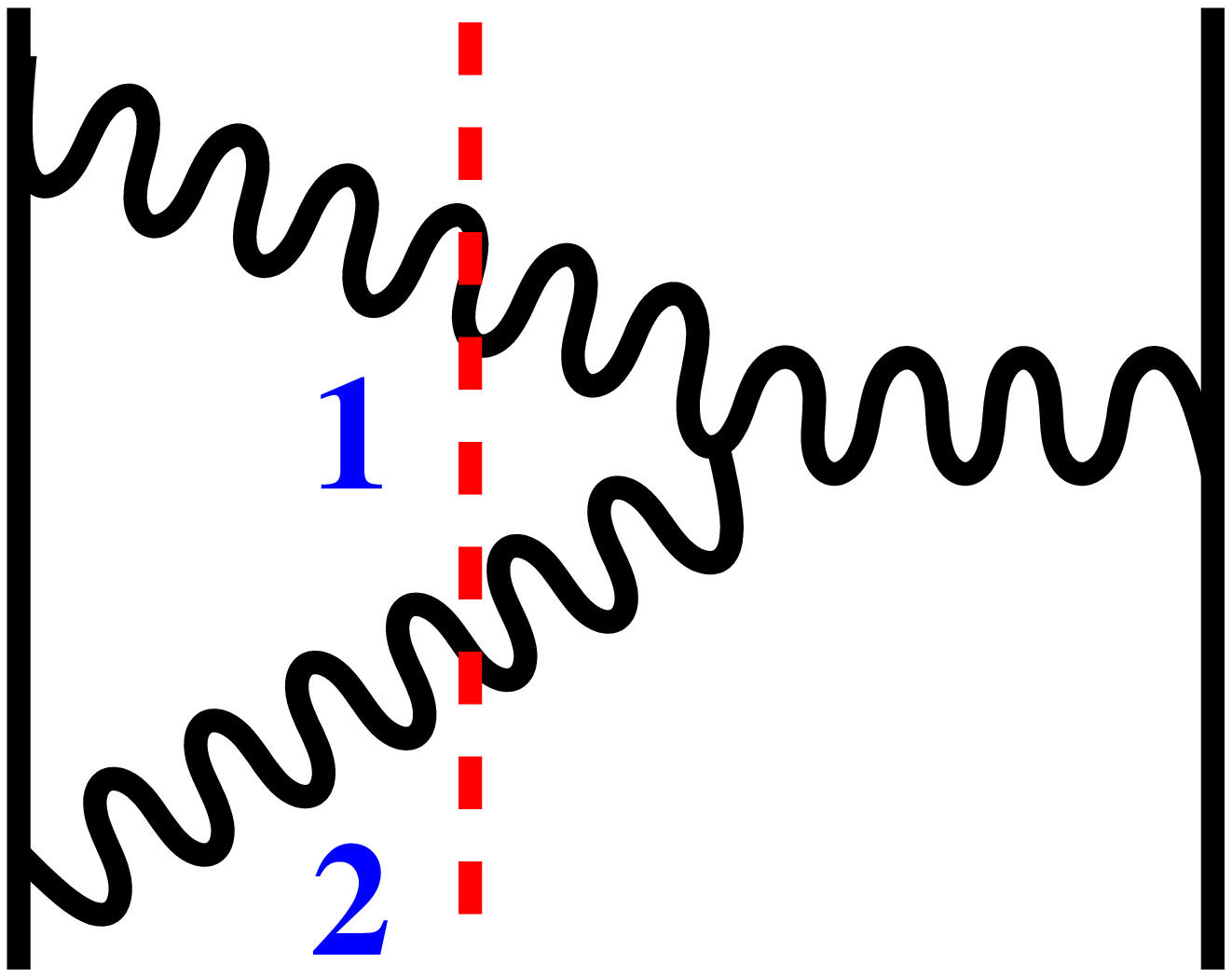, height = 20mm}
  \label{fig:nonabelc}}
\subfigure[]{
\epsfig{file=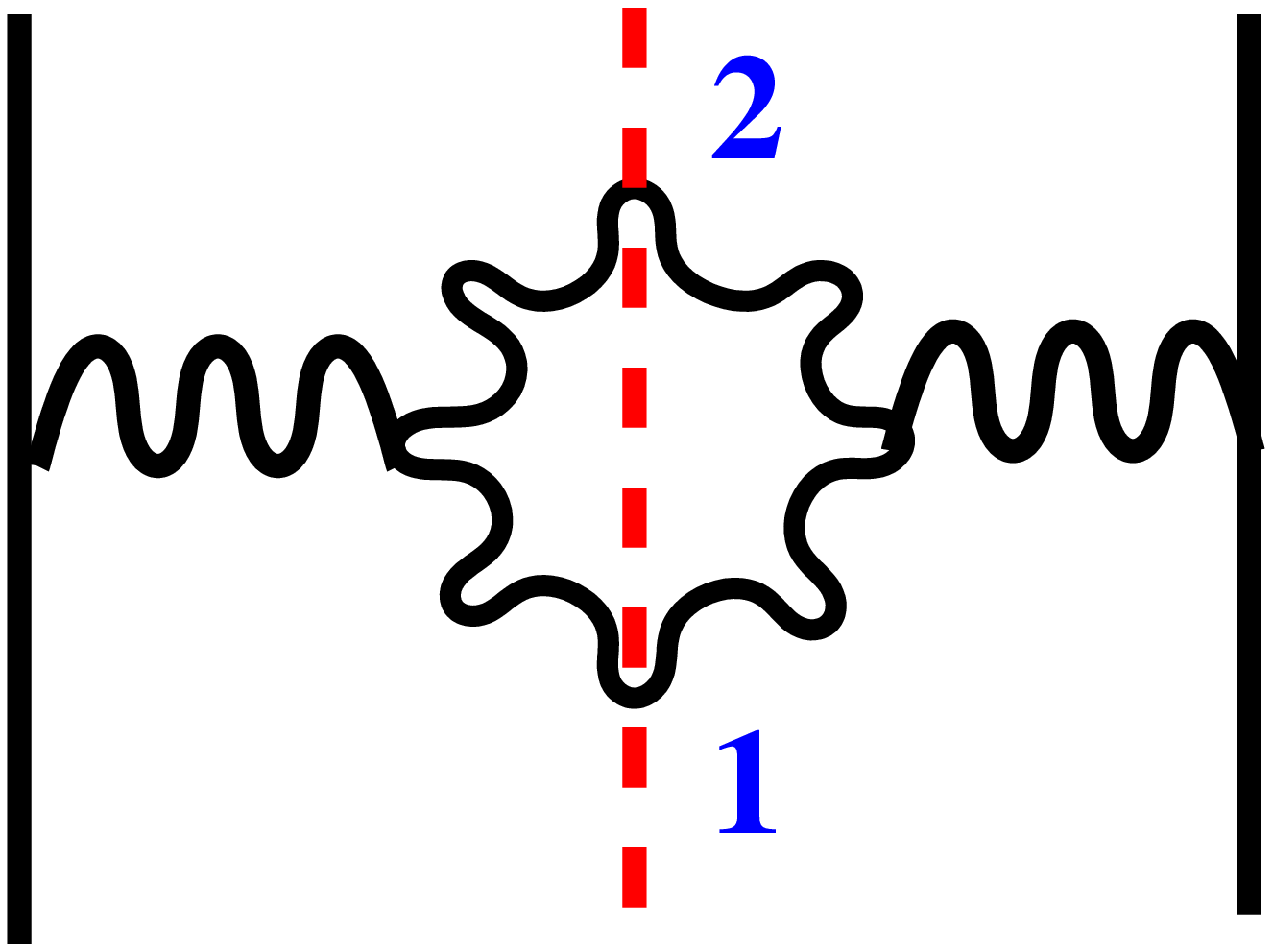,  height = 20mm}
 \label{fig:nonabeld}}
\caption{Two gluon Feynman diagrams contributing to non-singlet NLO kernels}
\label{two_glue_diags}
\end{centering}
\end{figure}

We shall analyze singular infra-red structure of the Feynman diagrams
depicted in Fig.~\ref{two_glue_diags}.
They describe emission of two gluons off a quark
and enter into calculation of the non-singlet NLO kernel.
Diagrams of Fig.~\ref{fig:laddera}-\ref{fig:ladderc}
are bremsstrahlung  type diagrams, including interference in Fig.~\ref{fig:ladderc}.
Their $C_F^2$ part is the same as in the corresponding case of QED,
hence we shall sometimes call them ``abelian''.
The other diagrams of Fig.~\ref{two_glue_diags}
include production of the gluon pair, see Fig.~\ref{fig:nonabeld},
and its interference with the previous bremsstrahlung diagrams,
Figs.~\ref{fig:nonabelb} and \ref{fig:nonabelc}.
Because of the presence of the triple-gluon vertex
they are diagrams of the genuine non-abelian origin.
Moreover, since the crossed-ladder diagram of Fig.\ref{fig:ladderc}
carries color factor equal to
$C_F^2-\frac{1}{2} C_FC_A $,
it contributes to both ``abelian'' and ``non-abelian'' part of NLO kernel.

Let us now introduce notation.
The two gluon phase-space is parametrized using Sudakov variables:
\begin{equation}
\begin{split}
&
k_i = \alpha_i p + \alpha_i ^- n + k_{i\perp}, \quad i = 1,2,
\\&
k = k_1 + k_2, \qquad q = p - k
\end{split}
\end{equation}
with $p$ being the four-momentum of the incoming quark
and $n$ a light-cone vector.
Four-vectors $k_1$ and $k_2$ denote four-momenta of the emitted gluons,
with their transverse parts being $k_{1\perp}$ and $k_{2\perp}$ respectively,
and $k^2=(k_1+k_2)^2$ being their effective mass.
The sum of  $\alpha_i$ is fixed
\begin{equation}
\alf1 + \alf2 = 1 - x.
\label{eq:x}
\end{equation}
We shall examine the distributions of two gluons in the soft limit:
\begin{equation}
\begin{cases}
\alf1 \rightarrow 0\cr
\alf2 \rightarrow 1-x
\end{cases}
\text{ or }
\begin{cases}
\alf1 \rightarrow 1-x\cr
\alf2 \rightarrow 0.
\end{cases}
\label{eq:sudakovlimit}
\end{equation}
We shall also use the ``eikonal minus variables'' $v_i$
of the emitted gluons defined as
\begin{equation}
\label{vi}
v_i = \frac{k_{i\perp}}{\sqrt{\alpha_i}}.%= \sqrt{2 pn}\sqrt{\alpha_1^-}
\end{equation}
The kernel is then extracted according to the scheme \cite{Curci:1980uw}
as a single pole in $\Gamma_0$:
\begin{equation}
\begin{split}
\Gamma_1  =
  \frac{C}{2\veps}\left(\frac{\alpha}{\pi}\right)^2
  \int d\Psi\delta(1- x -\alf1 - \alf2)
  \Theta(Q -\max\{v_{1}, v_{2}\} ) \rho(\alf1, \alf2, v_{1}, v_{2}, x),
\end{split}
\label{eq:gamma}
\end{equation}
where $C$ is the color factor and the function $\rho$
represents contribution from each Feynman diagram (trace and kinematics).
We shall also use the dimensionless ``eikonal phase-space'' defined as follows:
\begin{equation}
d\Psi =
 \frac{d\alf1}{\alf1} \frac{dv_1}{v_1} d\phi_1
 \frac{d\alf2}{\alf2} \frac{dv_2}{v_2} d\phi_2.
\label{Psi}
\end{equation}
The two-gluon phase space is cut from below
by means of geometrical regulator $\delta$, namely the factors $1/\alpha_i$ are regulated by principal value prescription: $\frac{1}{\alpha_i}\rightarrow\frac{\alpha_i}{\alpha_i^2+\delta^2}$.
The closing of the phase space from above is ensured by the $\Theta$ function\footnote{The choice of the variable in $\Theta$ function closing the phase space from above can be different. We use $\max\{v_1,v_2\}$ or $\max\{a_1,a_2\}$ which is different from CFP choice $q^2=-(p-k_1-k_2)^2$.}.
For the gluon pair mass we shall also use technical cut $k^2>\kappa$.
In the numerical exercises
we shall typically integrate \eqref{eq:gamma}
over the azimuthal angles $\phi_i$ of the gluons,
while concentrating on the dependence on $v_i$ and $\alpha_i$.
Also, because of the constraint in eq.~\eqref{eq:x},
if we say that we examine the distribution in $\alf1/\alf2$
it really means that we use $\alf1/(1-x-\alf1)$.
Also, due to a simple dimensional argument one of the variables $v_i$ can be always factored out from $\rho$ function and the essential dependence of the distributions
in variables $v_i$ can be reduced to
the dependence on the ratio $y=v_1/v_2$ only.

%%%%%%%%%%%%%%%%%%%%%%%%%%%%%%%%%%%%%%%%%%%%%%%%%%%%%%%%%%%%%%%%%%%
%%%%%%%%%%%%%%%%%%%%%%%%%%%%%%%%%%%%%%%%%%%%%%%%%%%%%%%%%%%%%%%%%%%
\section{IR cancellations among bremsstrahlung diagrams}
%%%%%%%%%%%%%%%%%%%%%%%%%%%%%%%%%%%%%%%%%%%%%%%%%%%%%%%%%%%%%%%%%%%
\label{sec:bremss}

In the first part we analyze "QED-like" bremsstrahlung diagrams of
Fig.~\ref{fig:laddera} - \ref{fig:ladderc}.
We shall first show infrared cancellations among these diagrams
in the  Monte Carlo exercise and later on analyze
the same cancellations analytically.

In the following numerical exercises we keep
variable $x$ fixed and equal 0.3.
This will ensure that at least one gluon is relatively hard.

The distributions
\begin{equation}
f(\alpha_i,v_i) = \int d\phi_1 d\phi_2 \delta(1-x-\alpha_1-\alpha_2) \delta\left(Q-\max\{v_1,v_2\}\right) \rho(\alpha_i,v_i)
\end{equation}
are plotted on Sudakov plane parametrized using variables
$\log(\alf1/ \alf2)$ and $\log(v_1/ v_2)$,
see eq.~\eqref{vi} for definition of $v_i$.

\begin{figure}[!h]
\epsfig{file=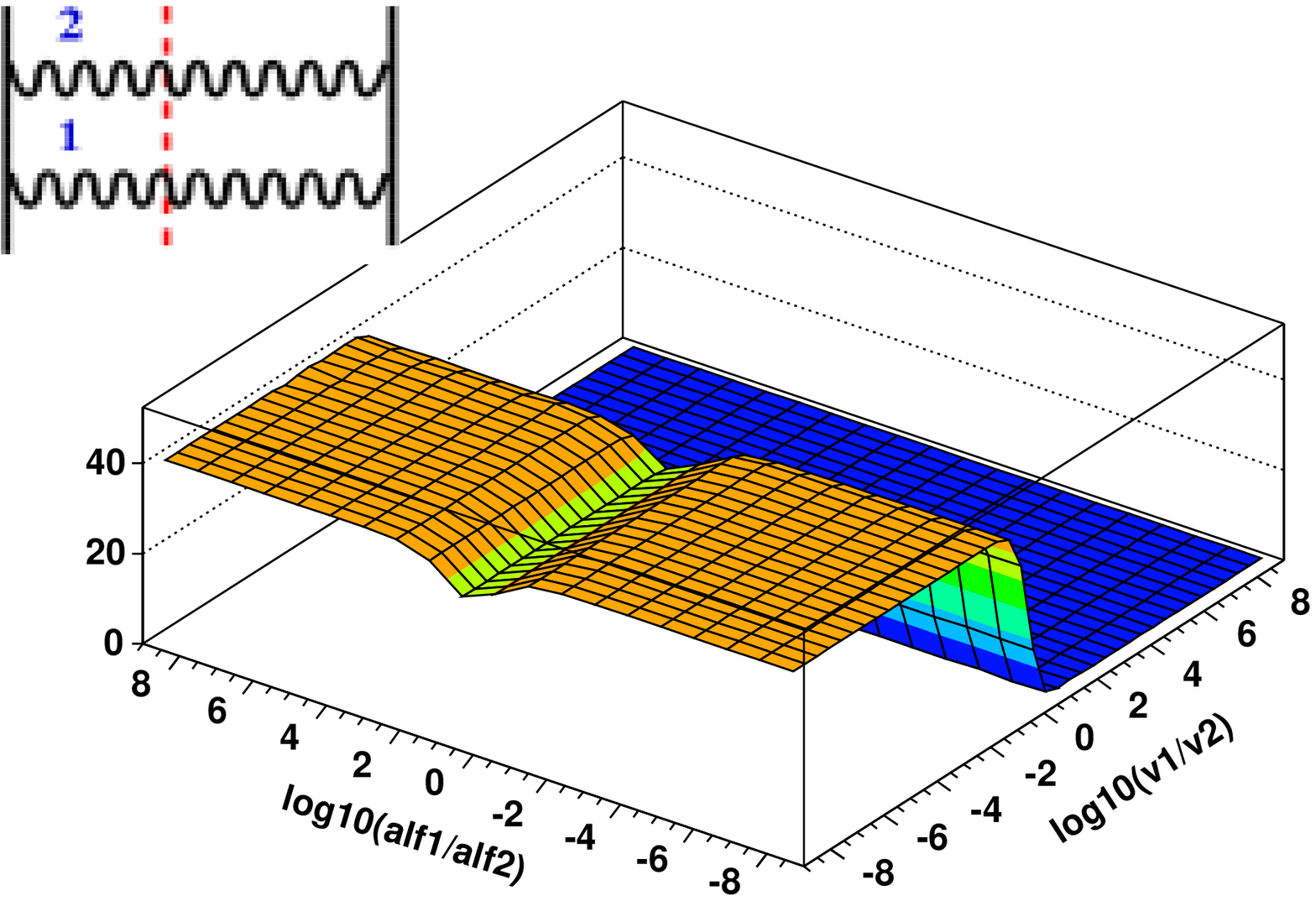, width=62mm}
\epsfig{file=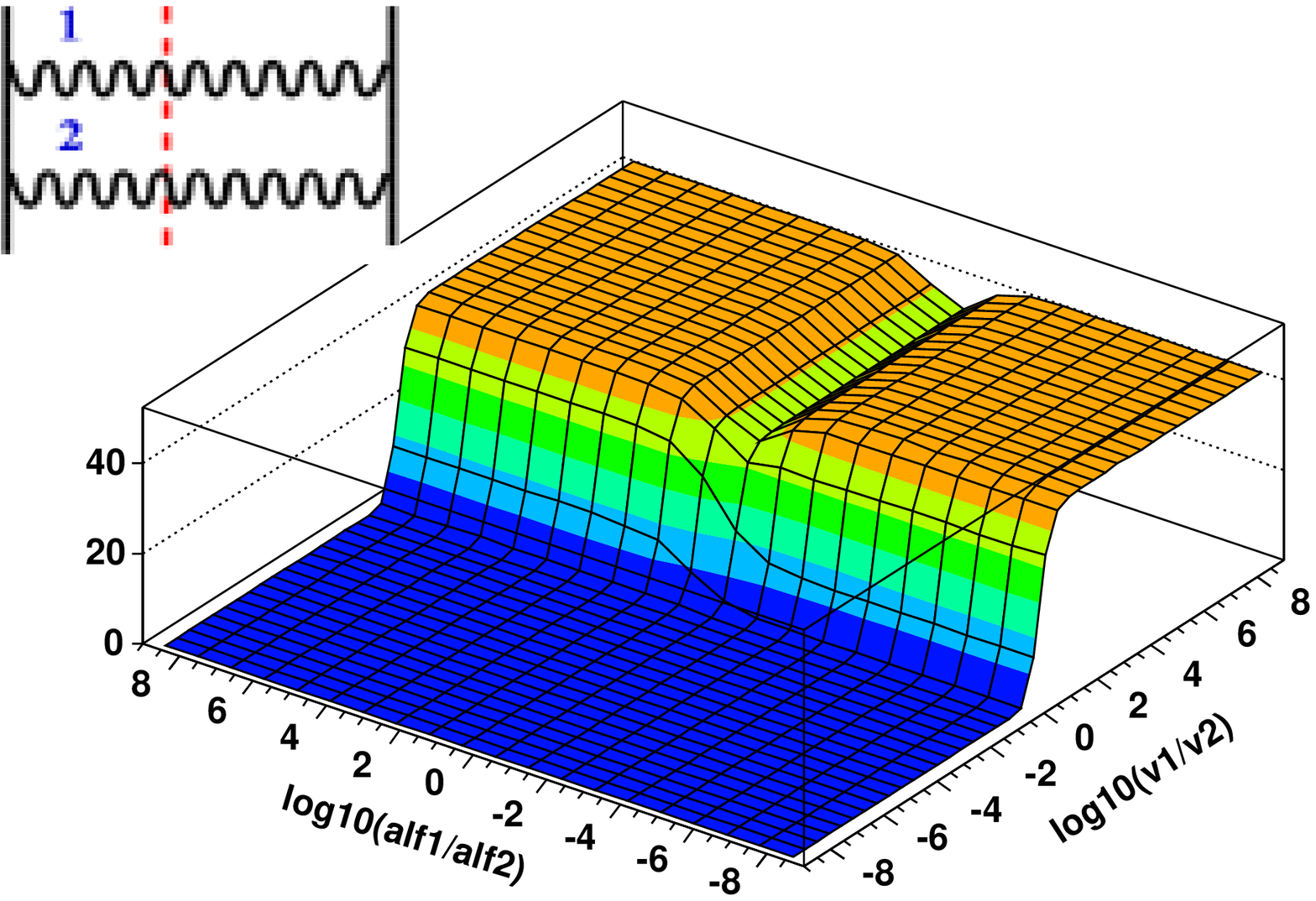, width=62mm}
\caption{Two double gluon emission ladders}
\label{nladders}
\end{figure}

\begin{figure}[!h]
\begin{centering}
\epsfig{file=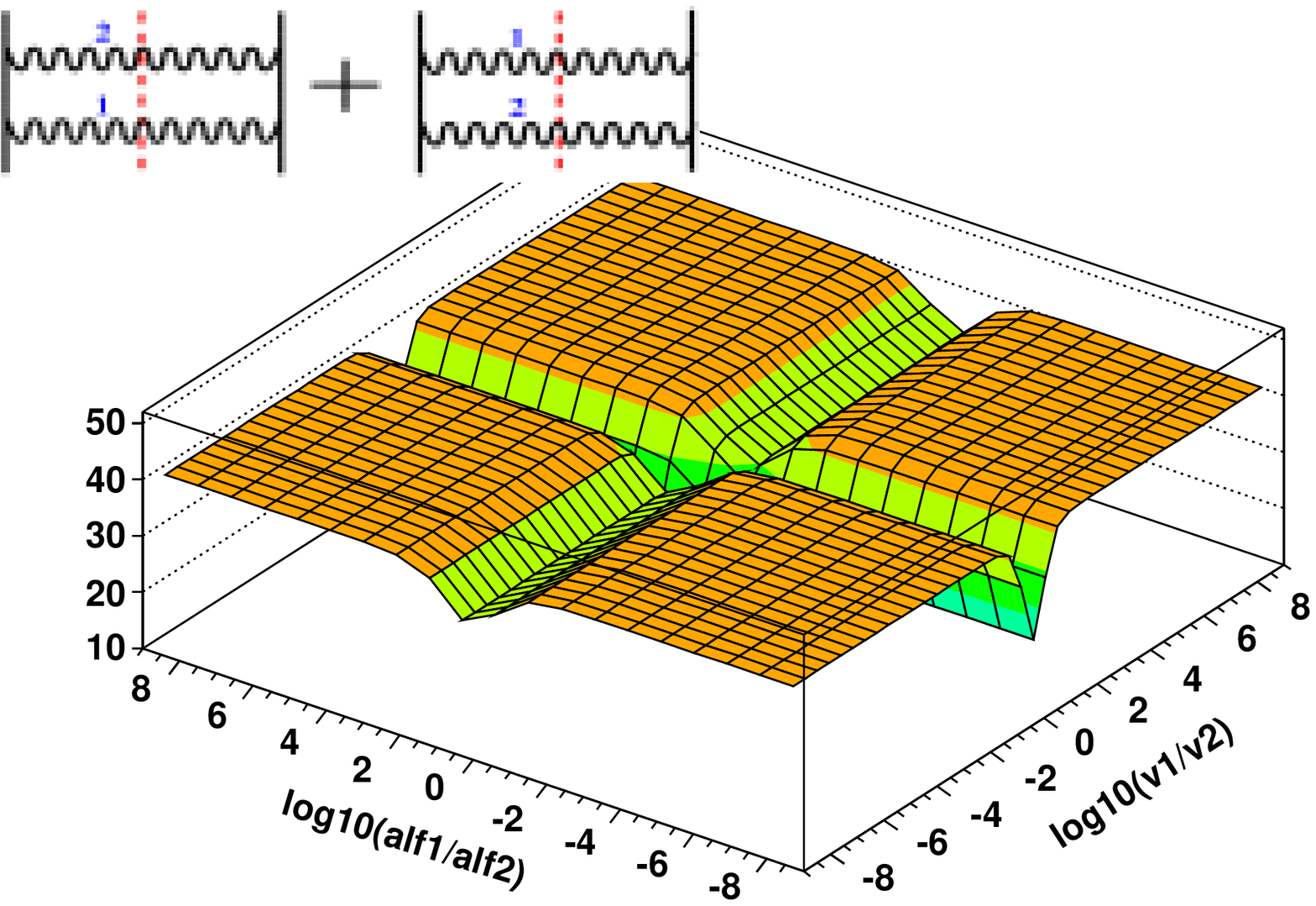,   width=62mm}
\epsfig{file=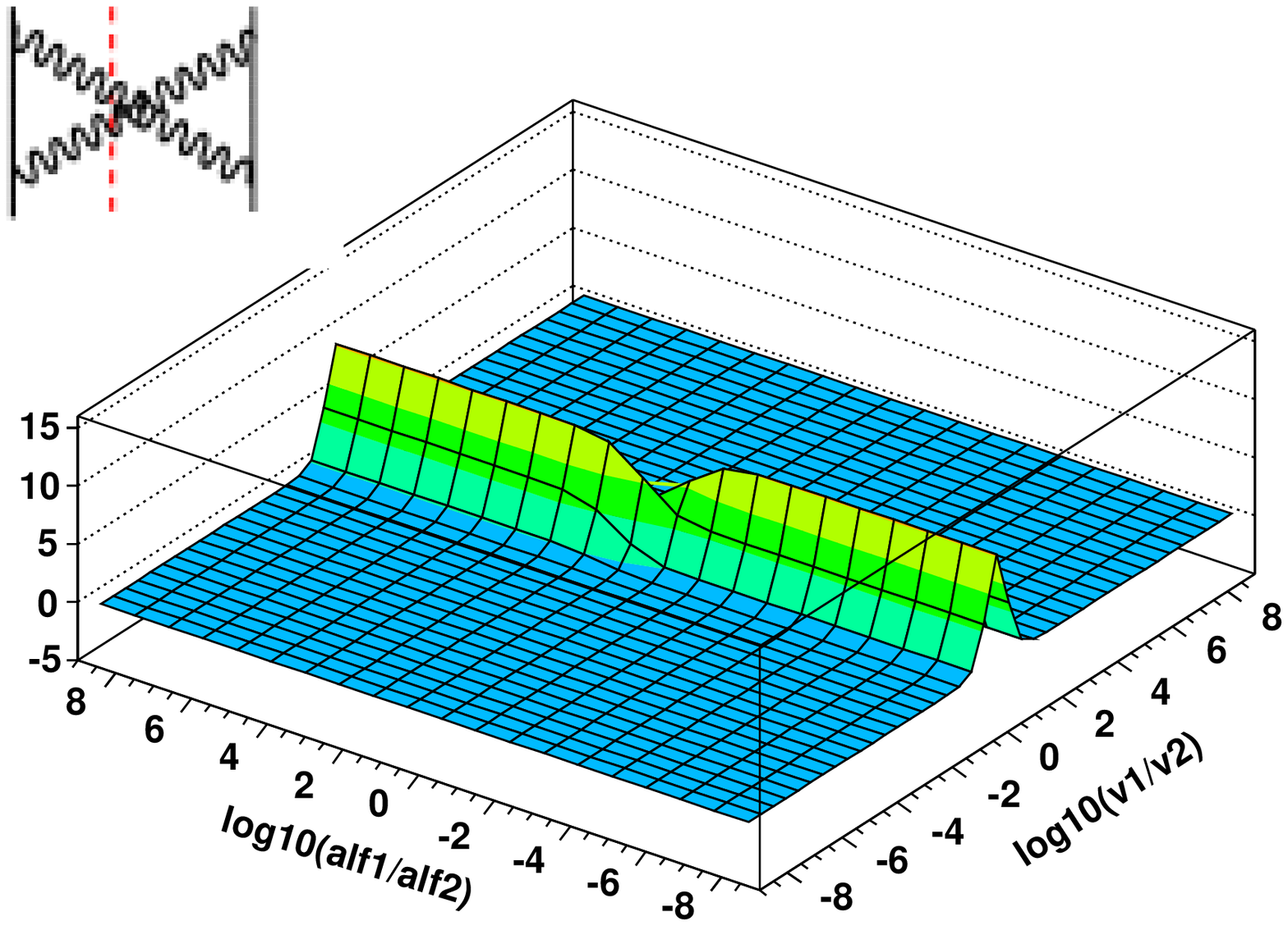,      width=62mm}
\epsfig{file=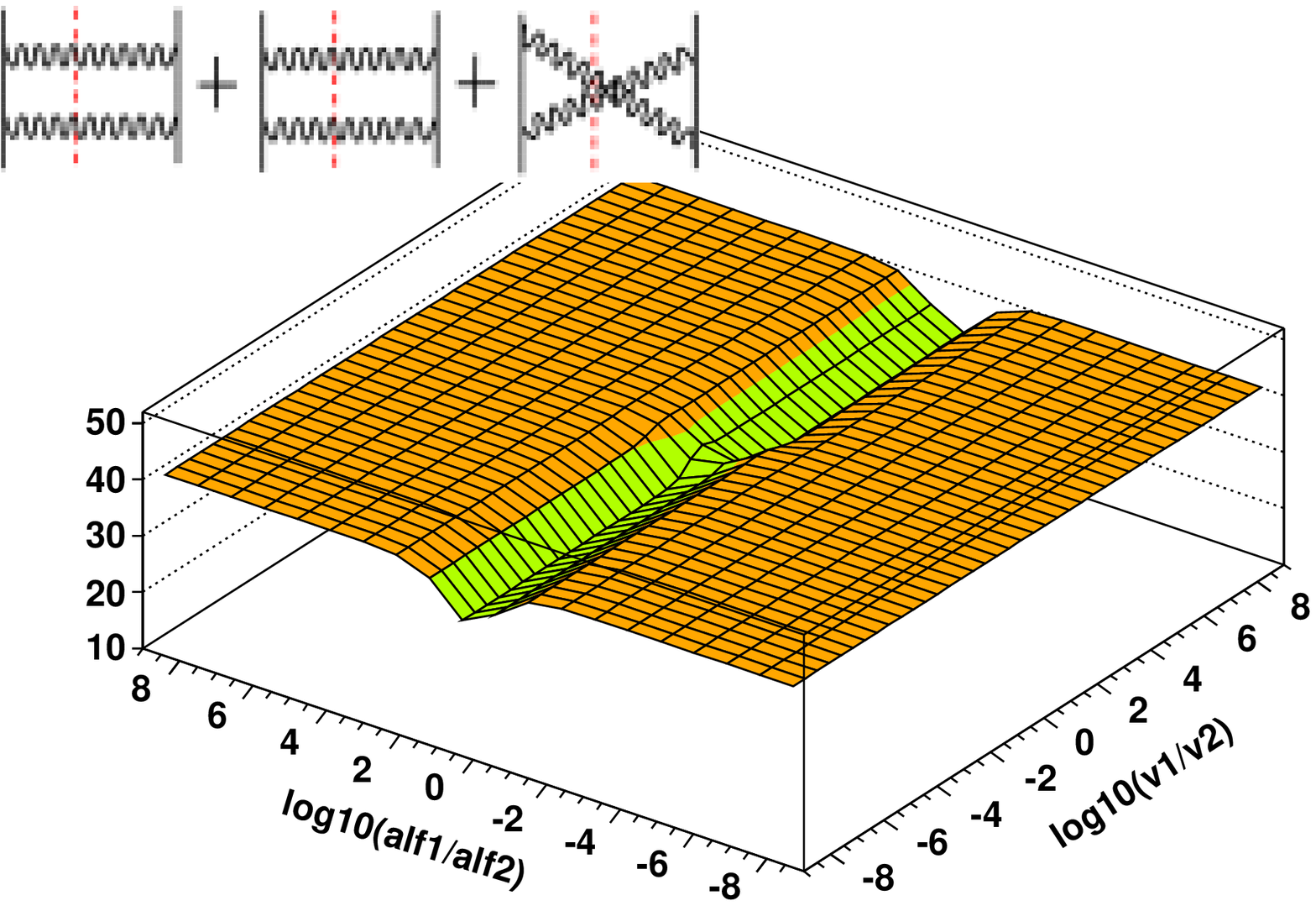, width=80mm}
\caption{
   Summarized bremsstrahlung diagram (left),
   crossed-ladder diagram (right) and their sum (down).
}
\label{nBrnBx}
\end{centering}
\end{figure}

Plots in Figs.~\ref{nladders}
show contributions from the two ladder diagrams.
They are obtained using Monte Carlo program FOAM \cite{foam:2002}.
As we see, their contributions appear to be strongly ordered
in virtuality variables $v_i$ of the emitted gluons,
see for instance left part of Fig.~\ref{nladders}.
The interference diagram is shown in the upper right plot of the Fig.~\ref{nBrnBx}.
It  contributes in the region of the phase-space
where both bremsstrahlung diagrams are comparable,
which is exactly the line of equal virtualities $v_1 = v_2$.
The crossed-ladder interference diagram has a singly-logarithmic singularity
along the same line $v_1 = v_2$.
%-----------------------------------------------------------------------------

%-----------------------------------------------------------------------------

%--------------------------------------------------------------

Fig.~\ref{nBrnBx} represents both contributions: from
summarized and crossed ladder (upper plots) together with their sum (lower plot).
The singly-logarithmic structure,
visible on the upper-left plot, disappears completely. 
The dominant contribution is an infinite ``plateau'',
with a long ``valley'' along the line
$\alf1 \simeq \alf2$,
This is not an internal infra-red singularity, however,
  but a leading-log structure. The ladder diagram is not 2PI,
  but quadratic in the leading-order kernel
  and therefore requires a soft counterterm to cancel this doubly-logarithmic plateau, see~\cite{ifjpan-iv-09-3}.
  This counterterm is necessary to construct the NLO kernel and its construction is justified by the CFP scheme. 
  We do not introduce it in this paper, as our aim is to present the universality of gauge cancellations. 
  Therefore we often refer to: ``contribution to the NLO kernel'' rather than a kernel itself, 
having in mind that this is not a complete NLO kernel in a strict sense.
Despite this dominant LO contribution,
in the Fig.~\ref{nBrnBx} one 
sees no structure in the soft Sudakov limit
~\eqref{eq:sudakovlimit}.

Results of the above numerical exercise can be also understood analytically.
Contributions from diagrams in  Figs.~\ref{fig:laddera} -~\ref{fig:ladderc}
in the soft limit are proportional to simple expressions shown in
Tab.~\ref{table:tab1}.
Two columns in this table refer to two possible different Sudakov limits,
with either first or second gluon being soft.

\begin{table}[!ht]
\center
\begin{tabular}[b]{|l|c|l|l|}
\hline
~~~ & ~~~~ & ~~~~& ~~~~ \\
    &~~~ &\Large $\alf1 \rightarrow 0$
    &\Large $\alf2 \rightarrow 0$ \\ 
~~~ & ~~~~ & ~~~~& ~~~~ \\
\hline
Br1 &   \epsfig{file=B1.eps, height = 11mm}
    &\Large $C_F^2 \frac{(1+x^2)}{(1+xy)^2}$ 
    &\Large  $C_F^2 \frac{(1+x^2)}{(y+x)^2}\; x^2$ \\ 
Br2 &   \epsfig{file=B2.eps, height = 11mm}
    &\Large  $C_F^2 \frac{(1+x^2)}{(1+xy)^2}\; x^2y^2 $ 
    &\Large  $C_F^2 \frac{(1+x^2)}{(y+x)^2}  y^2$\\
Bx  &   \epsfig{file=Bx.eps,   height = 11mm}
    &\Large $2C_F^2\frac{(1+x^2)}{(1+xy)^2}\; xy $
    &\Large  $2C_F^2\frac{(1+x^2)}{(y+x)^2}\; xy$ \\ 
\hline
~~~ & ~~~~ & ~~~~ &~~~~\\
& \Large{\bf SUM}
    &\large $C_F^2\left(1+x^2\right)$ 
    &\large $C_F^2\left(1+x^2\right)$ \\ 
~~~ & ~~~~ & ~~~~&~~~~\\
\hline
\end{tabular}
\caption{Contributions from ladders in the Sudakov limit (up to a constant factor) }
\label{table:tab1}%label must be after caption, for correct referencing
\end{table}

The common denominator
$ \frac{1}{(1+xy)^2}$
is the (rescaled) square of the virtual quark propagator
$\frac{1}{q^4}=\frac{1}{((p-k_1-k_2)^2)^2}$
after emitting two gluons.
Other factors come from $\gamma$-traces%
\footnote{Only ``$C_F^2$'' part of the crossed
  ladder is taken here, hence the color coefficient 
  is shown explicitly. Other unimportant factors are omitted.
  }.
The last row is the sum of the three, see also the lower plot in
Fig.~\ref{nBrnBx}. It is finite in the soft Sudakov limit,
as the denominator cancels out exactly with the spinorial part
of the matrix element squared.

In the above warm-up exercise we have examined in a fine detail how
in the NLO kernel calculations, in the axial gauge,
quantum interference cancellations do work in practice.
In fact they were exactly the same as in QED in axial gauge.
Let us now turn to the genuine non-abelian gauge cancellations
of the same kind.

%%%%%%%%%%%%%%%%%%%%%%%%%%%%%%%%%%%%%%%%%%%%%%%%%%%%%%%%%%%%%%%%%%%%%%%
%%%%%%%%%%%%%%%%%%%%%%%%%%%%%%%%%%%%%%%%%%%%%%%%%%%%%%%%%%%%%%%%%%%%%%%
\section{IR cancellations among genuine non-abelian contributions}
%%%%%%%%%%%%%%%%%%%%%%%%%%%%%%%%%%%%%%%%%%%%%%%%%%%%%%%%%%%%%%%%%%%%%%%

In this section we will discuss contributions
proportional to $C_F C_A$ from the diagrams (c-f) of Fig.~\ref{two_glue_diags}.
They are of the genuine non-abelian character, as certified by the presence of $C_A$.
It is interesting to see how they all ``communicate'' in the soft Sudakov limit
defined below.

We shall start with the overview of the IR cancellations
in analytical form and next we shall illustrate them with 2-dimensional
plots coming from Monte Carlo numerical exercises.
In the following
we find useful to use rapidity-related variables $a_i$
in order to parametrize phase-space of two gluons:
\begin{equation}
  a_i = \frac{k_{i\perp}}{\alpha_i},\qquad
  d\Psi =  \frac{d\alf1}{\alf1} \frac{da_1}{a_1} d\phi_1\,
           \frac{d\alf2}{\alf2} \frac{da_2}{a_2} d\phi_2.
  \label{eq:ai}
\end{equation}
The angular dependence enters only through the relative angle between $k_{1\perp}$ and $k_{2\perp}$ namely $\phi_{12}$, the remaining angle can be integrated out since nothing depends on it.

By the soft Sudakov limit we understand that
\begin{equation}
\begin{cases}
\alpha_i \rightarrow 0\cr
k_{i\perp} \rightarrow 0
\end{cases}
\label{eq:softsudakov}
\end{equation}
while $a_i$ is finite. 
The Sudakov plane in the plots will be parametrized using variables
$\log(\alf1/\alf2)$ and $\log(a_1/a_2)$.

%------------------------------------------------------------------
\begin{table}[h!]
\center
\begin{tabular}[t]{|l|l|l|}
\hline
Vg & \epsfig{file=Vg.eps,  height = 10mm, width=13mm}
& \large $C_F C_A (1+x^2)\left[\; \frac{u^2}{a^2} F_1
              + \; \frac{1}{a^2} F_2 \;\right]$
\\
Yg1 & \epsfig{file=Yg1.eps, height = 10mm, width=13mm}
& \large $C_F C_A (1+x^2)\left[\; \frac{u\cos\phi_{12} -u^2}{a^2} F_1
          +x\; \frac{\alpha_2}{1-x} F_0\;\right]$
\\
Yg2 & \epsfig{file=Yg2.eps, height = 10mm, width=13mm}
& \large $C_F C_A (1+x^2)\left[\; \frac{u\cos\phi_{12} -1}{a^2} F_2
          +x\; \frac{\alpha_1}{1-x} F_0\;\right]$
\\
Bx & \epsfig{file=Bx.eps,  height = 10mm, width=13mm}
& \large  $-C_F C_A (1+x^2)\; x F_0$
\\
\hline
~~~ & ~~~~ & ~~\\
& \Large{\bf SUM}
& \large $C_FC_A (1+x^2)\; \frac{u\cos\phi_{12}}{a^2}\left(F_1+F_2\right) $  \\
~~~ & ~~~~ & ~~\\
\hline
\end{tabular}
\caption{
Cancellations of the non-abelian contributions from various diagrams
due to gauge invariance (color coherence).
Singular part for each diagram is shown in analytical form.
Factors $F_i$ are relatively mild, see text for their definition.
}
\label{table:tab9}
\end{table}

%%%%%%%%%%%%%%%%%%%%%%%%%%%%%%%%%%%%%%%%%%%%%%%%%%%%%%%%%%%%%%%%%%%%%%%%%
\subsection{General structure of the IR non-abelian cancellations}

In Table~\ref{table:tab9} we summarize the IR cancellations of the non-abelian
origin between the two real gluon diagrams contributions to NLO kernel
in analytical form.
Formulas in Tab.~\ref{table:tab9}
are the leading divergences extracted
from two real gluon distributions in a maximally simplified form.
Functions $F_i$ are relatively mild and defined as follows:
\begin{equation}
\begin{split}
&F_1=\frac{\alf2}{\alf1}\frac{1}{u^2}F_0,\qquad
F_2=\frac{\alf1}{\alf2}u^2F_0,
\\&
\sqrt{F_0}=\frac{\sqrt{\alf1\alf2}u}%
  {\alpha_1(1-\alpha_2)u^2+\alpha_2(1-\alpha_1) +2\alpha_1\alpha_2 u\cos\phi_{12}}
= \frac{\sqrt{\alpha_1 \alpha_2}a_1 a_2}{q^2},
\end{split}
\end{equation}
where we defined $u = a_1/a_2$.
Moreover we define $a^2 = 1 + u^2 - 2u\cos\phi_{12}$,
which is up to the $\alpha_1\alpha_2a_2^2$ factor,
the effective mass of the gluon pair squared $k^2=(k_1+k_2)^2$.
Function $F_0 $ is proportional to rescaled  square of the propagator $1/q^4$, which is regular in the soft limit.

The pattern of the IR cancellations in Tab.~\ref{table:tab9} is manifest.

The most evident singularity is associated with $u/a^2$ factor, 
infinite when the effective mass of the gluon pair is zero. 
In logarithmic variables this $u/a^2$ divergence is seen as a thin infinite ridge along $u=1$. 
It is a function of $u$ and $\phi_{12}$ only, strongly peaked at $u=1$.
In the  gluonic vacuum polarization diagram this singularity is dominant. 
This is, however, a collinear  singularity, not a soft one. 
It  remains uncancelled but it is not relevant in our discussion of cancellation of 
soft singularities. We only mention it to understand the full singularity structure 
of the diagrams of interest.
In Yg1 and Yg2  diagrams the  $u/a^2$ factor is present, too but the singularity 
is cancelled out  by the numerators. 
% and therefore will not be discussed

The factors $u F_1$ and $F_2/u$ are functions of both $u$ and 
$\alpha_1 / \alpha_2$. They ``soften'' the sharp fall of $u/a^2$ 
in the limits $u\rightarrow 0$ and $u\rightarrow \infty$. They are non-zero in the soft Sudakov limit \eqref{eq:softsudakov}, giving rise to a doubly-logarithmic IR divergence%
\footnote{The doubly-logarithmic IR divergence after phase space integration is $\sim\ln^2\frac{1}{\delta}$, where $\delta$ is cut-off variable, see \em{Introduction}.}.
These terms are present in both Vg and Yg diagrams, with opposite signs.
This is checked and discussed below in the context of numerical exercises.

In the last row of Tab.~\ref{table:tab9} the sum of all aforementioned contributions 
is presented. The terms proportional to $u^2F_1/a^2$ and $F_2/a^2$ cancel 
explicitly among diagram with gluonic vacuum polarization and its
 interference with bremsstrahlung. 
The remaining factor $F_1+F_2$ is equal to 1 in the soft Sudakov limit \eqref{eq:softsudakov}, leaving out $u/a^2$ factor free from doubly-logarithmic divergences.

Let us stress that
this particular cancellation of the doubly logarithmic
Sudakov structure of the non-abelian origin
in the two-gluon distribution in QCD is usually referred to in the literature
as the ``color coherence effect'', see for instance ref.~\cite{khoze-book}.

The other IR divergence is caused by the presence of terms $\sim F_0$ 
in diagrams Yg1, Yg2 and Bx, see again Table~\ref{table:tab9}.  
$F_0$ is nonzero along $a_1^2/a_2^2 = \alpha_1/\alpha_2$, 
giving rise to a single-log singularity after phase space integration. 
This has been discussed already in the  case of bremsstrahlung diagrams.
Here, however,  the analytical cancellation among Yg1, Yg2 and Bx, 
ensured by $\alpha_1+\alpha_2=1-x$, occurs in the whole phase-space, 
not only in the soft limit. 

Finally, the only singularities that remain are associated with $u/a^2$ term,
as discussed before in this section.

%%%%%%%%%%%%%%%%%%%%%%%%%%%%%%%%%%%%%%%%%%%%%%%%%%%%%%%
\begin{figure}
\begin{centering}
\epsfig{file=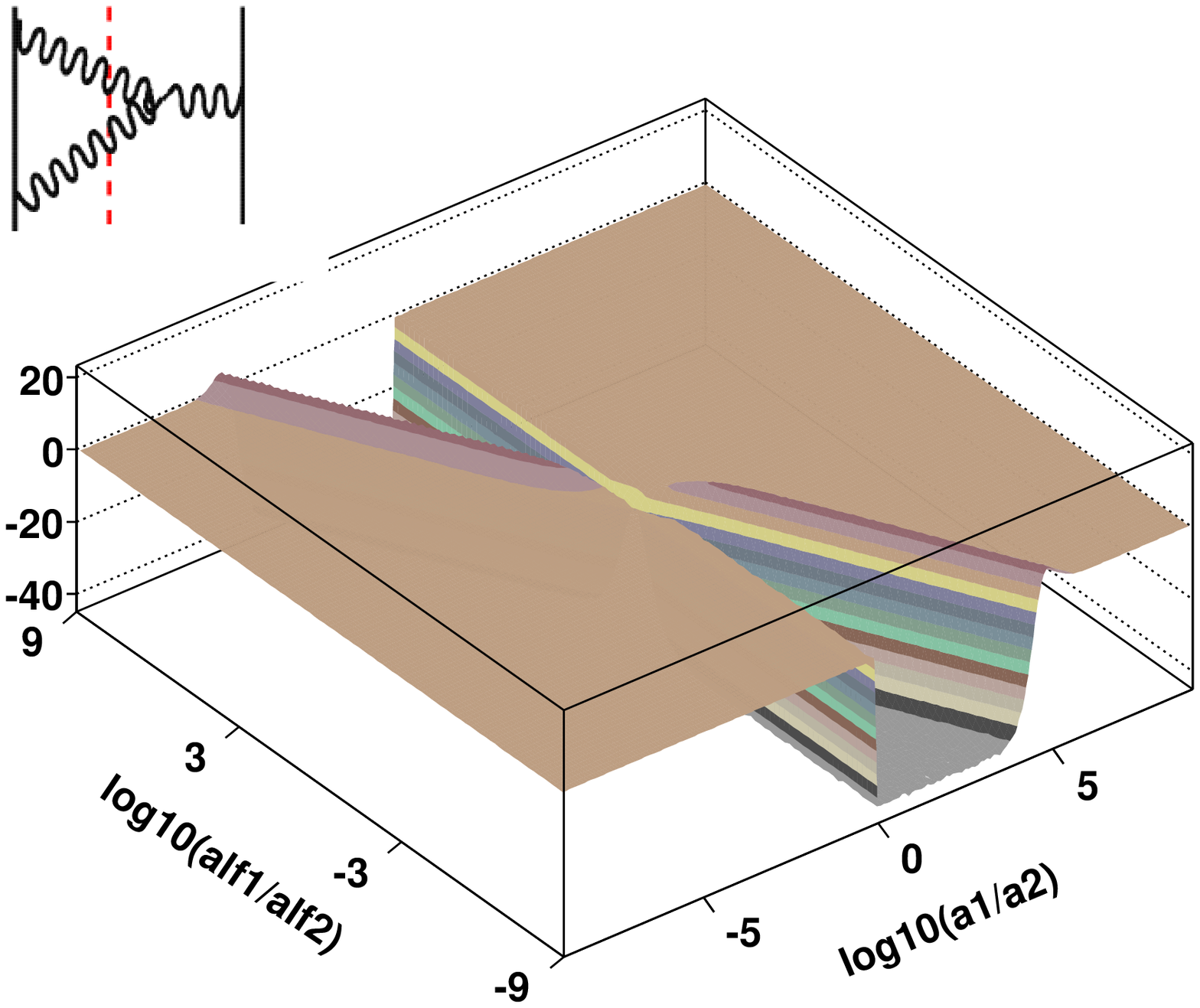, width=62mm}
\epsfig{file=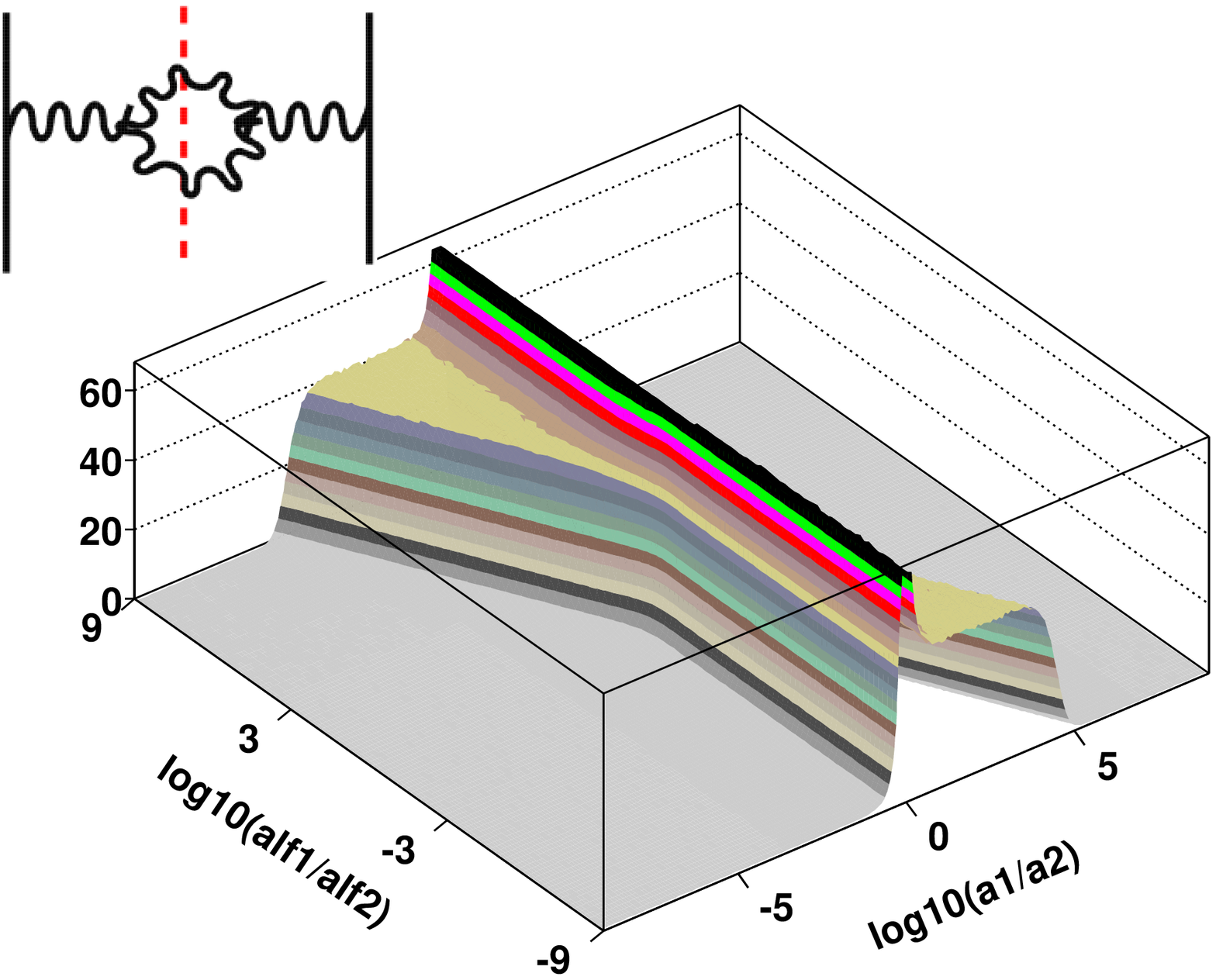, width=62mm}
\caption{Triple-gluon-vertex diagrams.}
\label{fig:rapi}
\end{centering}
\end{figure}

%%%%%%%%%%%%%%%%%%%%%%%%%%%%%%%%%%%%%%%%%%%%%%%%%%%%%%%%%%%%%%%%%%%%%%%%%
\subsection{Numerical illustration of IR non-abelian cancellations}

\begin{figure}
\begin{centering}
\epsfig{file=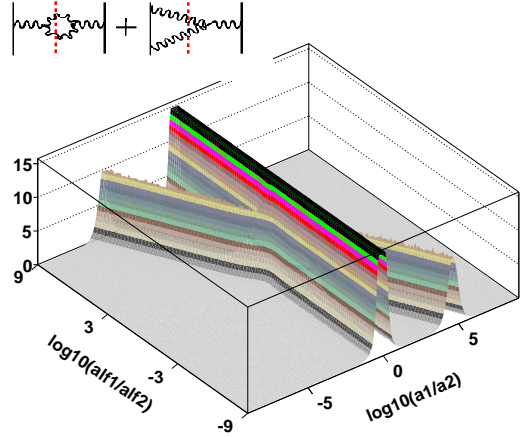, width=80mm}
\caption{Full cancellation of double-logs between the vacuum polarization diagram and bremsstrahlung-vacuum polarization interference}
\label{fig:Vg+Yg_rapi}
\end{centering}
\end{figure}

\begin{figure}[h!]
\begin{centering}
\epsfig{file=eVg+YgRapi.eps, width=62mm}
\epsfig{file=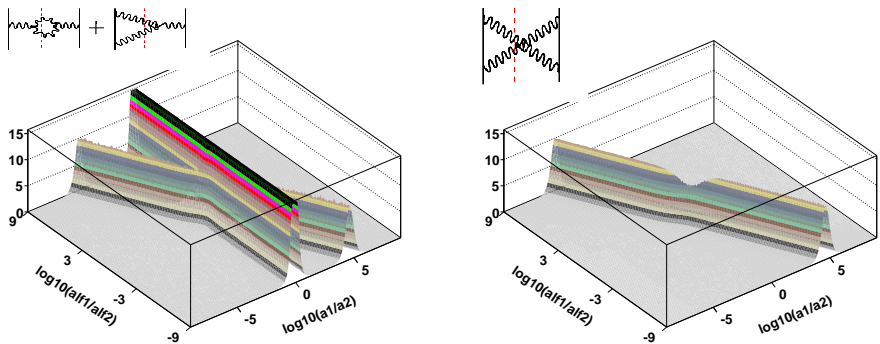,    width=62mm}
\caption{The previous plot (left) and bremsstrahlung diagram (right)}
\label{two_I0}
\end{centering}
\end{figure}

\begin{figure}[h!]
\begin{centering}
\epsfig{file=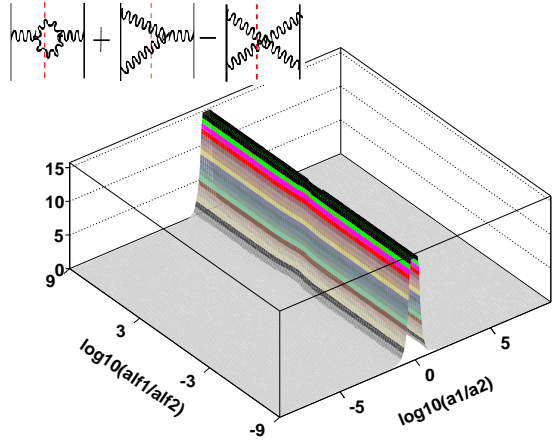, width=8cm}%, height = 5cm}
\caption{$C_FC_A$ part of the sum (Vg+Yg $-$ Bx). }
\label{VYall}
\end{centering}
\end{figure}

In Fig.~\ref{fig:rapi} we show the distribution $f(a_i,\alpha_i)$
for two gluons (averaged over the gluon azimuthal angles)
for fixed $x=1-\alpha_1-\alpha_2 = 0.3$,
from gluonic pair production graph (Vg)
and its interference with bremsstrahlung (Yg = Yg1+Yg2),
see also Tab.~\ref{table:tab9}. Contributions from all diagrams are written explicitly in Tab.~\ref{table:tab9}. 
In the plots on  Figs.~\ref{fig:rapi}~-~\ref{VYall} we omit their color coefficients.
The gluonic pair production graph (Vg) has a strong peak along the line
of equal rapiditites $a_1=a_2$ originating from $u/a^2$ factor (off-shell gluon propagator).
In addition, this diagram features in the plot triangular infinite plateau
between the lines of the equal rapiditites $a_1=a_2$
and the line of equal virtualities $v_1=v_2$.
It is exactly this triangular plateau which upon integration leads to $\ln^2(1/\delta)$.
Note, that in Figs.~\ref{fig:rapi}~-~\ref{VYall} we use variables $\log (a_1/a_2)$, contrary to the previous section, where we had $\log (v_1/v_2)$. This is why the line $v_1=v_2$ is now diagonal in the plots.

Very similar, but with opposite sign, doubly-logarithmic structure 
is present in the left plot in Fig.~\ref{fig:rapi}
from the interference graphs Yg1+Yg2, see also Tab.~\ref{table:tab9}.
After adding the contributions from the above diagrams,
see Fig.~\ref{fig:Vg+Yg_rapi}, the doubly-logarithmic structure between
the lines: $a_1 = a_2$ and $v_1=v_2$ disappears.
What remains, is the single-log singularity,
appearing in a familiar shape along the diagonal line of $v_1=v_2$
(barely visible in the plots of Fig.~\ref{fig:rapi}) 
and collinear singularity represented here as an infinite ridge along the line of equal rapiditites. 
The latter is associated with zero effective mass of the gluon pair, $k^2$.

%%%%%%%%%%%%%%%%%%%%%%%%%%%%%%%%%%%%%%%%%%%%%%%%%%%%%%%
%%%%%%%%%%%%%%%%%%%%%%%%%%%%%%%%%%%%%%%%%%%%%%%

The above cancellation is the well known ``color coherence effect'',
see for instance ref.~\cite{khoze-book}.
It reflects the fact that the gluonic pair production graph (Vg)
pretends that in the triangular region it is the harder gluon
carrying octet color charge which is the ``emitter'',
with the emission strength $\sim C_F C_A$.
In reality, however, the emitter should be quark carrying triplet color charge,
with the almost twice weaker emission strength $\sim C_F^2$.
The role of the interference diagram Yg1+Yg2 is to correct for that
and we see this to happen in the plot
and in the formulas in Tab.~\ref{table:tab9}.
One has to remember that this part of the plot is already populated with the
bremsstrahlung diagrams of the previous section proportional to $\sim C_F^2$.

What still remains in Fig.~\ref{fig:Vg+Yg_rapi}
is a singly-logarithmic structure along the diagonal line $v_1=v_2$.
Its presence is in principle allowed in the doubly-logarithmic Sudakov approximation.
A more subtle analysis of the soft limit in QCD shows that it should
also vanish and this phenomenon is often referred to as ``eikonalization'',
see for instance ref.~\cite{yfs:1961,Frenkel:1983da}.
The job of bringing back the proper soft limit of the two gluon distribution
and eliminating remaining single-log structure is done by the crossed
bremsstrahlung diagram Bx (in fact its $C_FC_A$ part), as it is shown
in Figs.~\ref{two_I0} and \ref{VYall}, see also Tab.~\ref{table:tab9}.

In right hand side of Fig.~\ref{two_I0}
the crossed-ladder diagram Bx is presented again.
On the left hand side of Fig.~\ref{two_I0}
we show the result of adding the triple-gluon vertex diagrams (Vg+Yg1+Yg2).
Both plots have a characteristic single-log structure,
seen as an infinite ridge along the diagonal line of equal virtualities $v_1=v_2$.
However, Bx  has opposite signs in its $C_F C_A$ color coefficients.
In Fig.~\ref{VYall} we see the result of adding all the 
above `` non-abelian'' diagrams. 
Bx enters with a minus sign.
We see that the singly-logarithmic
structure $\sim C_F C_A$  disappears completely in the soft limit
$\alpha_i\to 0$.
What is still present  in the picture is the dominant $1/k^2 \sim u/a^2$
gluon pair mass singularity.

%%%%%%%%%%%%%%%%%%%%%%%%%%%%%%%%%%%%%%%%%%%%%%%

%%%%%%%%%%%%%%%%%%%%%%%%%%%%%%%%%%%%%%%%%%%%%%%
%%%%%%%%%%%%%%%%%%%%%%%%%%%%%%%%%%%%%%%%%%%%%%%

\begin{figure}
\begin{centering}
\epsfig{file=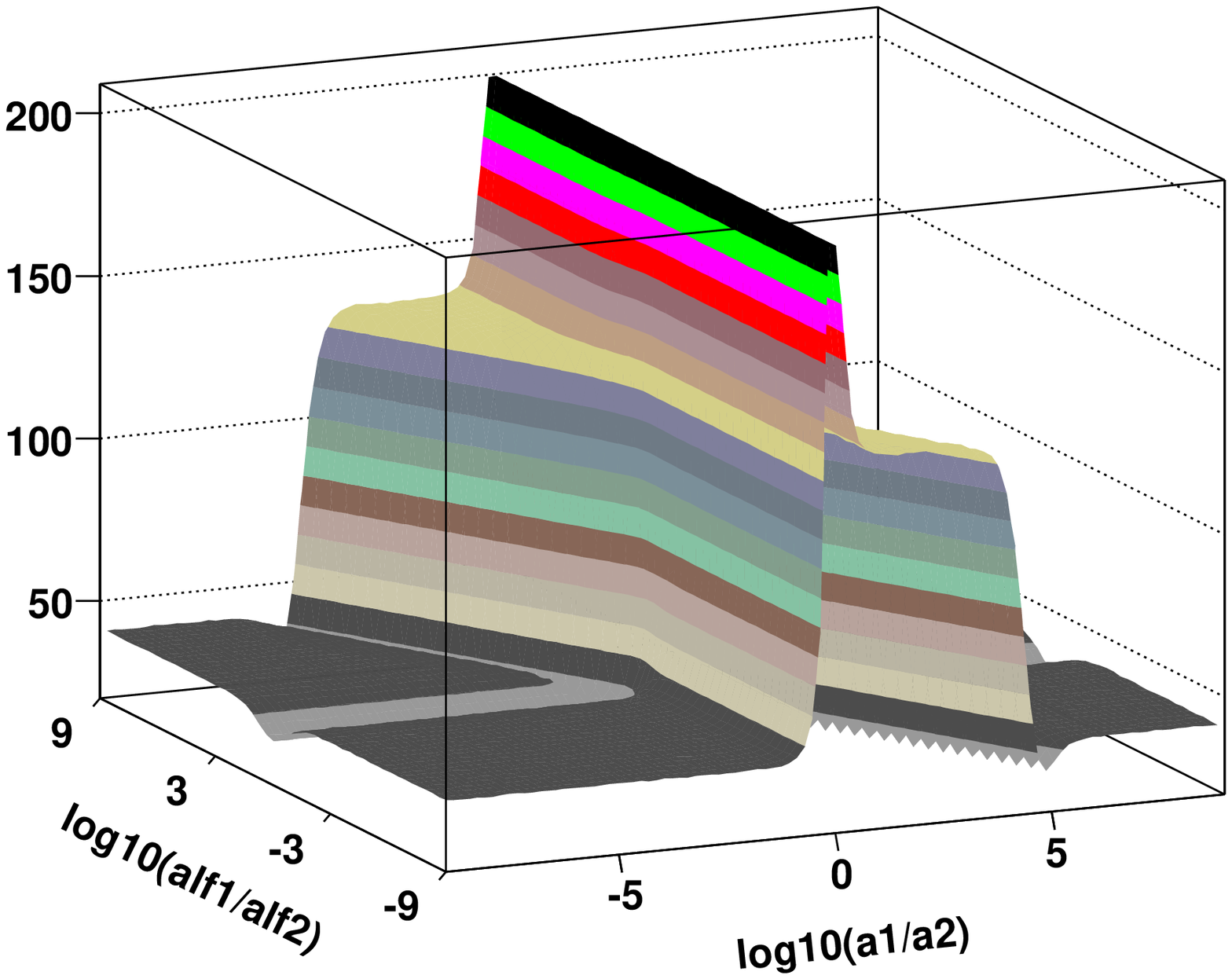, width=60mm}
\epsfig{file=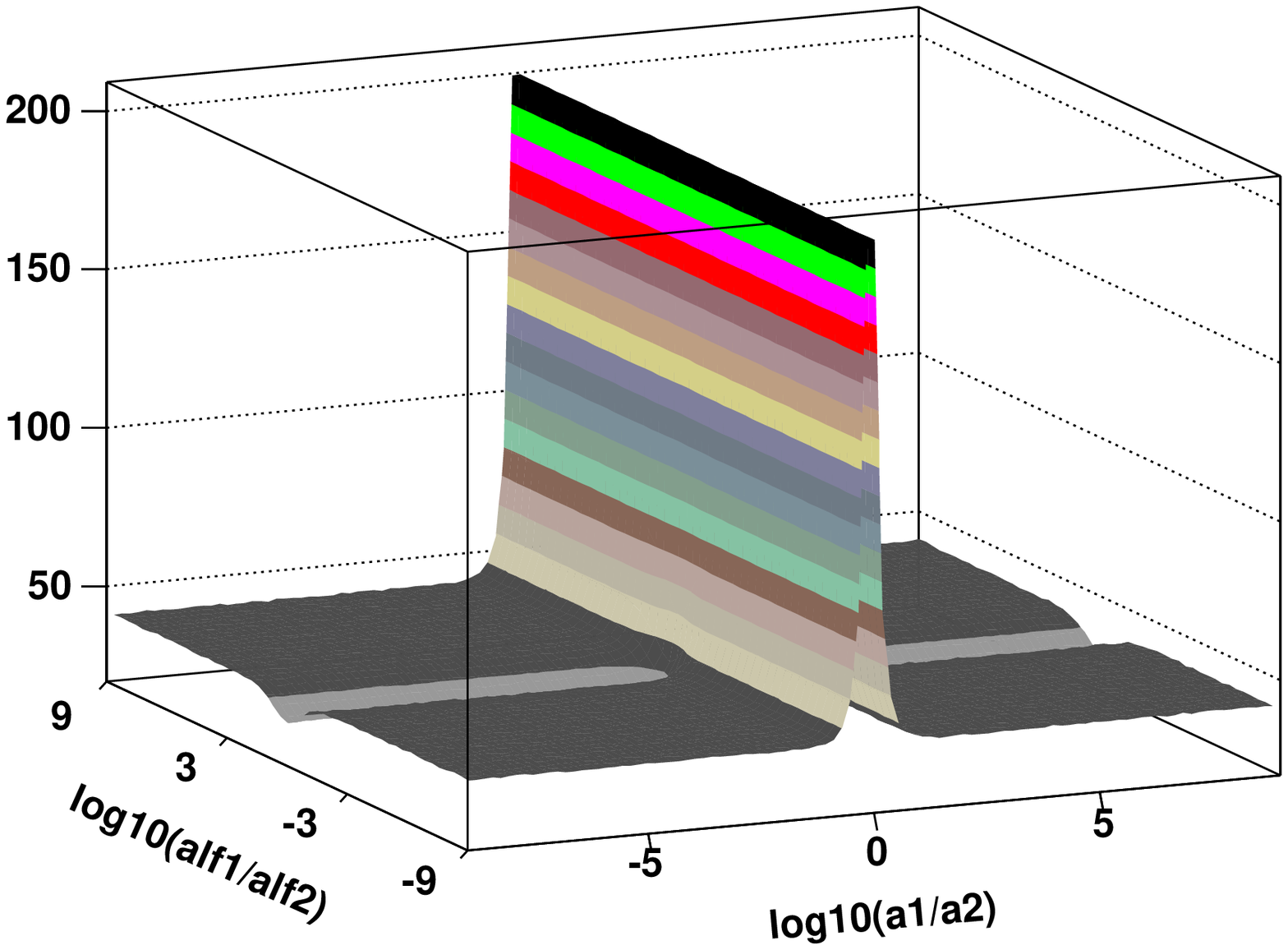, width=60mm}
\caption{Infra-red cancellations among NLO non-singlet diagrams. Br1, Br2 and Vg (left), Br1, Br2, Bx, Vg, Yg1 and Yg2 (right). }
\label{fig:All_rapi}
\end{centering}
\end{figure}

The concluding plots are shown in Fig.~\ref{fig:All_rapi}.
We presented there Feynman diagrams entering the kernel, both
$\sim C_F^2$ and $\sim C_FC_A$.
In the left plot there are
solely amplitude-squared diagrams (Br1, Br2 and Vg)  
and in the right plot all diagrams including interferences.
We see explicitly the crucial role of ``color coherence effects'' (the interference diagrams) in the cancellation of IR singularities.
In the sum of all diagrams of interest we see that remaining structure lies on top of the LO doubly-logarithmic plateau. 
The plateau does not enter into the NLO kernel.  It is cancelled by the counterterm required by the kernel definition, 
as discussed in Section 2.

%%%%%%%%%%%%%%%%%%%%%%%%%%%%%%%%%%%%%%%%%%%%%%%%%%%%%%%

%%%%%%%%%%%%%%%%%%%%%%%%%%%%%%%%%%%%%%%%%%%%%%%%%%%%%%%%%%%%%%%%%%%%%%%
\section{Conclusions}

We examined the infra-red structure of the diagrams contributing to 
NLO  non-singlet kernel in the unintegrated form.

We have shown the mechanisms of gauge cancellations occurring among different diagrams 
and the importance of ``color coherence effects'' for this cancellations.

These effects in soft Sudakov limit are examined/discussed in both
analytical and numerical form.

\vspace{4mm}
\noindent
{\bf Acknowledgments}\\
We would like to thank Stanis{\l}aw Jadach, Maciej Skrzypek  and Boris Ermolaev for many useful discussions during the preparation of this work. 

%%%%%%%%%%%%%%%%%%%%%%%%%%%%%%%%%%%%%%%%%%%%%%%%%%%%%%%%%%%%%%%%%%%%%%%%%%%%
%\bibliographystyle{utphys_spires}
%\bibliographystyle{h-physrev3}
%\bibliography{radcor}

\begingroup\begin{thebibliography}{1}

\bibitem{DGLAP}
L.N. Lipatov, {\em Sov. J. Nucl. Phys.} {\bf 20} (1975) 95;\\ V.N. Gribov and
  L.N. Lipatov, {\em Sov. J. Nucl. Phys.} {\bf 15} (1972) 438;\\ G. Altarelli
  and G. Parisi, {\em Nucl. Phys.} {\bf 126} (1977) 298;\\ Yu. L. Dokshitzer,
  {\em Sov. Phys. JETP} {\bf 46} (1977) 64.

\bibitem{ifjpan-iv-09-3}
S.~Jadach and M.~Skrzypek, report IFJPAN-IV-09-3, to appear in these
  proceedings.
\href{http://arxiv.org/abs/0905.1399}{{\tt arXiv:0905.1399  [hep-ph]}}.

\bibitem{Curci:1980uw}
G.~Curci, W.~Furmanski, and R.~Petronzio, {\em Nucl. Phys.} {\bf B175} (1980)
27.
%%CITATION = NUPHA,B175,27;%%.

\bibitem{Ellis:1978sf}
R.~K. Ellis, H.~Georgi, M.~Machacek, H.~D. Politzer, and G.~G. Ross, {\em Phys.
  Lett.} {\bf B78} (1978)
281.
%%CITATION = PHLTA,B78,281;%%.

\bibitem{foam:2002}
S.~Jadach, {\em Comput. Phys. Commun.} {\bf 152} (2003) 55--100,
\href{http://www.arXiv.org/abs/physics/0203033}{{\tt physics/0203033}}.
%%CITATION = PHYSICS 0203033;%%.

\bibitem{khoze-book}
Y.~Dokshitzer, V.~Khoze, A.~Mueller, and S.~Troyan, {\em Basics of Perturbative
  QCD}.
\newblock Editions Frontieres, 1991.

\bibitem{yfs:1961}
D.~R. Yennie, S.~Frautschi, and H.~Suura, {\em Ann. Phys. (NY)} {\bf 13} (1961)
  379.

\bibitem{Frenkel:1983da}
J.~Frenkel, J.~G.~M. Gatheral, and J.~C. Taylor, {\em Nucl. Phys.} {\bf B228}
  (1983)
529.
%%CITATION = NUPHA,B228,529;%%.

\end{thebibliography}\endgroup
\providecommand{\href}[2]{#2}
%%%%%%%%%%%%%%%%%%%%%%%%%%%%%%%%%%%%%%%%%%%%%%%%%%%%%%%%%%%%%%%%%%%%%%%%%%%

\end{document}